\documentclass[11pt,twoside]{article}
\usepackage{asp2010}

\aspcpryear{2011}
\aspvoltitle{ADASS XX (to appear)}
\aspvolauthor{Evans, I.~N. and Accomazzi,~A. and Mink, D.~J. and Rots, A.~H.}

\resetcounters

\begin{document}

\title{Digital Preservation and Astronomy: Lessons for funders and the funded}
\author{Norman Gray and Graham Woan
\affil{School of Physics and Astronomy, University of Glasgow, UK}}

\begin{abstract}
Astronomy looks after its data better than most disciplines, and it is
no coincidence that the consensus standard for the archival preservation
of all types of digital assets -- the OAIS Reference Model -- emerged
originally from the space science community.

It is useful to highlight both what is different about astronomy (and indeed
about Big Science in general), what could be improved, and what is
exemplary, and in the process I will give a brief introduction to the
framework of the OAIS model, and its useful conceptual vocabulary.  I
will illustrate this with a discussion of the spectrum of big-science
data management practices from astronomy, through gravitational wave (GW)
data, to particle physics.


\end{abstract}

\section{Introduction}

In this paper we will briefly discuss some of the issues surrounding
the `long-term' management of data for `big science' projects.  It is therefore
intended to illuminate the ways in which scientists in these areas
have distinctive data management requirements, and a distinctive data
culture, which contrasts informatively with other disciplines.

Below, we give a brief account of what we mean by `big science',
illustrate some
of the features of astronomy, particle-physics and gravitational wave
data, present an overview of the OAIS model, and summarise the work we have
done in this area.


\section{Big data in big science}

What is `big science'?

Big science projects tend to share many features which distinguish
them from the way that experimental science has worked in the past.
Such projects share (non-independent) features such as:
\begin{description}
\item[big money] These are decades-long projects, supported by
  country-scale funders and billion-currency-unit budgets;
\item[big author lists] The (gravitational wave) LSC
  author list runs at around 0.8\,kAuthor, and the LHC's ATLAS detector
  is 3\,kAuthor;
\item[big data rates] Projects produce petabytes of data per year (ATLAS will flatten out at
  about 10\,PB/yr, and Advanced LIGO at around 1\,PB/yr);
\item[big administration] These include MOUs, councils, and more.
\end{description}
There is an interesting historical discussion of the features of `big science',
and LIGO's progress towards that style of working, in~\cite{collins03}, with an
extended history of the sub-discipline in~\cite{collins04}.  The study
we are reporting on is restricted to the case of such large-scale projects.

What is the `long term'?

Astronomy has a long tradition of preserving data, of exploiting old
data, and also, in the teeth of centuries of technological change, of
finding old data to be readable.  The `Venus of Ammisaduqa' cuneiform
tablet (a 7th century BCE media refresh of 17th century observations)
is intelligible only to antiquaries, but the 12th century Toledan
tables are intelligible, with help, to a contemporary astronomer, and
Kepler's 17th century Rudolphine tables are readable without
difficulty, suggesting that the intelligibility timescale for
astronomy is of order one millennium (note that each of these examples
are disseminated publications, not raw data).  Particle physics is
more mayfly-like, with HEP data becoming unreadable and uninteresting
on a timescale of a few deades.  Gravitational Wave astronomy is
somewhere in the middle, with an experimental practice reminiscent of
particle physics, but (eventual) data products which match astronomy.

Data preservation requires a consideration of both media preservation
and data intelligibility on timescales potentially ranging from months
(in the case of researchers wanting to re-find their own data) to
multiple millennia (in the case of very long term nuclear waste
disposal \citep{wipp04}).  The Open Archival Information System (OAIS)
standard, however, pragmatically defines the `long-term' to mean, essentially, `long
enough for storage technology to change'.

\begin{figure}
\begin{centering}
\includegraphics[scale=1.25]{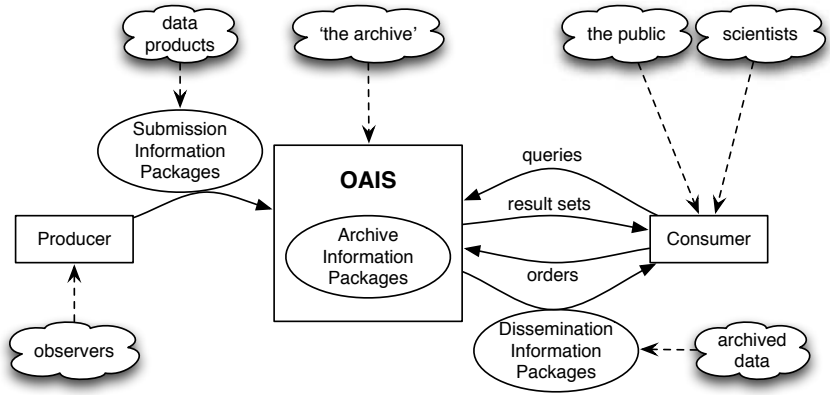}\\
\end{centering}
\caption{\label{f:oais}A high-level view of the OAIS model, with
  annotations indicating the non-OAIS names for objects (redrawn from
  \citet{std:oais})}
\end{figure}

The OAIS standard \citep{std:oais} is a high-level model for archives
(see Fig.~\ref{f:oais}).  Since the OAIS model ultimately emerged from
the space-data community, it is a natural fit to astronomical data.
Although the OAIS model has been criticised for being excessively
general -- it is hard to think of any half-way respectable archive
that is not at some level `conformant' with OAIS \citep[cf.,][]{rosenthal05} --
it is nonetheless very useful as a conceptual toolbox, and as the
basis for costings and other deliberations.

Once data sets have been deposited in an archive, and thus reprocessed
from Submission Information Packages (SIP) to Archive ones (AIP), they become the
exclusive responsibility of the archive, which must plan, and be
funded, accordingly.  One of the key questions to be decided by the
archive is the nature of the Designated Communities who are expected
to retrieve information from the archive, and the description of these
communities -- who they are, what they know, how hard they can be
expected to work in order to retrieve archived data -- is a key part
of the archive's documentation.  That documentation is written with
enough detail that anyone in the designated communities can use it to
make the data intelligible, \emph{without} recourse to any of the data
producers, who are presumed to have died, retired, or forgotten all
they know, and the creation of this `Representation Information' is a
key part of the archive's initial negotiation with the data producers.




\section{Sharing and preserving astronomy data}

In science, we preserve data so that we can make it available
later. This is on the grounds that scientific data should generally be
universally available, partly because it is usually publicly paid for,
but also because the public display of corroborating evidence has been
part of science ever since the modern notion of science began to
emerge in the 17th century (CE) -- witness the Royal Society's motto,
which loosely translates as `take nobody's word for it'.  Of course,
the practice is not quite as simple as the principle, and a host of
issues, ranging across the technical, political, social and personal,
make this more complicated than it might be, but it is worth noting
that the physical sciences generally perform better here than other
disciplines, both in the technical maturity of the existing archives
and in the community's willingness to allocate the time and money to
see this done effectively.

In 2009--2010 the authors were commissioned by JISC, which is
responsible for the exploitation of digital technology in
the UK HE system, to examine the way in which the gravitational wave
community, as a proxy for big science in general, managed its data,
and to make recommendations as appropriate to JISC, the relevant
funding councils, and to the GW community.  The results of this work
will appear at the URL \texttt{http://purl.org/nxg/projects/mrd-gw} in
due course.  The brief summary of the conclusions is that this
community, accurately representing the broader astronomical community,
is functioning very well in this respect.

In particular, the commendable features of the community's approach to
data management are:
\begin{itemize}
\item There is an expectation of explicit and costed data management
  planning (though this could arguably benefit from being more systematic).
\item There is implicit identification of `designated communities'.
\item There has always been a recognition that clearly available and
  described  science-data products (AIPs in OAIS-speak) are a vital
  part of the communication between observers and their colleagues.
\item The community is honest about both its need for proprietory
  periods, and the acknowledgement that these restrictions must only
  be temporary and short.
\end{itemize}
All of these features are naturally re-expressible in terms of the
OAIS model, which conveniently couples them to the work being done
elsewhere on the costing of OAIS models, and on developing tests of
OAIS conformance.

Our project has not yet come to specific conclusions about funding.
However, it becomes clear that funders are, at some level, as
concerned with predictability and auditability as with economy, so
that if conventional practice can be demonstrated to be additionally
good practice, or within easy reach of it, then this facet of funders'
goals can be marked as achieved.

Finally, we believe that our recommendations will resemble the following.
\begin{enumerate}
\item Big-science funders should require projects to develop plans
  based on the OAIS model, or profiles of it;
\item they should additionally develop or support expertise in
  criticising the result;
\item and use that modelled plan as a framework for validation of the
  project's efforts, both during the lifetime of the project and at
  its end.
\end{enumerate}

\textbf{Glossary}
\emph{ATLAS}: one of the detectors at the LHC;
\emph{LIGO}: the US-based gravitational wave experiment;
\emph{LSC}: the multinational LIGO Scientific Collaboration;

\acknowledgements
This talk describes work at the University of Glasgow, funded by JISC, UK.
We are grateful for helpful discussions with the LSC;
LIGO document P1000179-v1.

\bibliography{O6-3}
\bibliographystyle{asp2010}

\end{document}